\documentclass[preprint,aps,prl,floatfix]{revtex4-2}

\usepackage[pdftex]{graphicx}
\usepackage{amsmath}
\usepackage{ragged2e}

\begin{document}

\title{Odd-parity electronic order near the semiconductor limit}

\author{Jack Tregidga}
\affiliation{Materials Department, University of California, Santa Barbara, California, USA}

\author{Dibyata Rout}
\affiliation{Materials Department, University of California, Santa Barbara, California, USA}

\author{Johannes Hielscher}
\affiliation{Materials Department, University of California, Santa Barbara, California, USA}

\author{Josiah Turner}
\affiliation{Materials Department, University of California, Santa Barbara, California, USA}

\author{Stephen D. Wilson}
\affiliation{Materials Department, University of California, Santa Barbara, California, USA}

\author{John W. Harter}
\email[Corresponding author: ]{harter@ucsb.edu}
\affiliation{Materials Department, University of California, Santa Barbara, California, USA}

\date{\today}

\begin{abstract}
Identifying materials platforms in which dilute carriers experience strong Coulomb interactions is a central challenge in the search for interaction-driven quantum phases. In such systems, weak carrier screening can promote a variety of collective instabilities beyond the conventional Fermi liquid paradigm, including superconductivity, Wigner crystallization, and odd-parity electronic order. Experimental realizations of such dilute, strongly interacting electronic systems remain rare in crystalline materials. Here we report a spontaneous odd-parity phase transition in the phosphide semiconductor family \textit{Ln}Cd$_3$P$_3$ (\textit{Ln} = La, Ce, Pr, Nd). Using optical second harmonic generation, we observe the onset of bulk inversion and rotational symmetry breaking accompanied by the emergence of an in-plane polar axis. Second harmonic microscopy reveals three domain variants related by 120$^\circ$ rotations, while ultrafast transient reflectivity measurements uncover a pronounced electronic reconstruction across the transition. Remarkably, the ordered phase appears only in lightly self-hole-doped compounds and is absent in insulating SmCd$_3$P$_3$, indicating an essential role for itinerant carriers despite their extremely low concentration. Guided by density functional theory, we develop a four-band model of the valence states and show that modest interactions can stabilize odd-parity electronic order. The resulting phase combines a spontaneous Fermi surface distortion with a momentum-dependent bilayer polarization that breaks inversion symmetry. Our results establish a route to interaction-driven parity breaking in dilute-carrier semiconductors and identify honeycomb bilayer systems as a promising platform for odd-parity electronic phases.
\end{abstract}

\maketitle

Electronic instabilities that spontaneously lower crystal symmetry are a central theme in quantum materials science, particularly in systems where reduced dimensionality, orbital degeneracy, and spin-orbit coupling cooperate to amplify interaction effects. Among these, odd-parity electronic phases that break inversion symmetry without relying on conventional lattice-driven ferroelectricity have emerged as an especially intriguing frontier because they enable unconventional transport and optical responses, spin textures, and intertwined orders~\cite{shi2013,fu2015,harter2017,bhowal2023,nagaosa2024,rodriguez2025,li2026}. To date, however, most candidate inversion-breaking electronic phases have been identified either in correlated metals with substantial carrier densities or in materials with preexisting structural tendencies toward polar distortions. Whether a purely electronic instability can stabilize a polar phase in a lightly doped semiconducting system with only a dilute population of itinerant carriers remains largely unexplored. In particular, the possibility of a spontaneous Fermi surface distortion that simultaneously breaks rotational and inversion symmetry has remained experimentally elusive.

Here we investigate the layered phosphides \textit{Ln}Cd$_3$P$_3$ (\textit{Ln} = lanthanide), a family of quasi-two-dimensional semiconductors whose low-energy electronic structure derives predominantly from degenerate $p_x$ and $p_y$ orbitals on weakly coupled CdP honeycomb layers~\cite{nientiedt1999,higuchi2016,feng2019,chamorro2023,lee2019,chamorro2025,alvarado2025,jang2025}. In LaCd$_3$P$_3$, temperature-dependent optical second harmonic generation (SHG) measurements reveal an odd-parity phase transition at $T_c = 190$~K accompanied by spontaneous rotational symmetry breaking. SHG rotational anisotropy (RA-SHG) establishes the emergence of an in-plane polar axis, while SHG microscopy uncovers three symmetry-related domain variants. Remarkably, we find the odd-parity phase occurs only in samples that are intrinsically self-hole-doped. Motivated by these observations, we develop a tight-binding model derived from density functional theory (DFT) and show within mean-field theory that modest electronic interactions can stabilize an odd-parity phase even at low carrier densities. Our results identify lightly doped semiconductors as a previously unexplored setting for interaction-driven polar metallicity and establish \textit{Ln}Cd$_3$P$_3$ as a platform in which dilute itinerant carriers can induce a collective electronic instability with odd parity.

\section{Results}

\subsection{Odd-parity phase transition in LaCd$_3$P$_3$}

\begin{figure*}[t]
\includegraphics[width=6.4in]{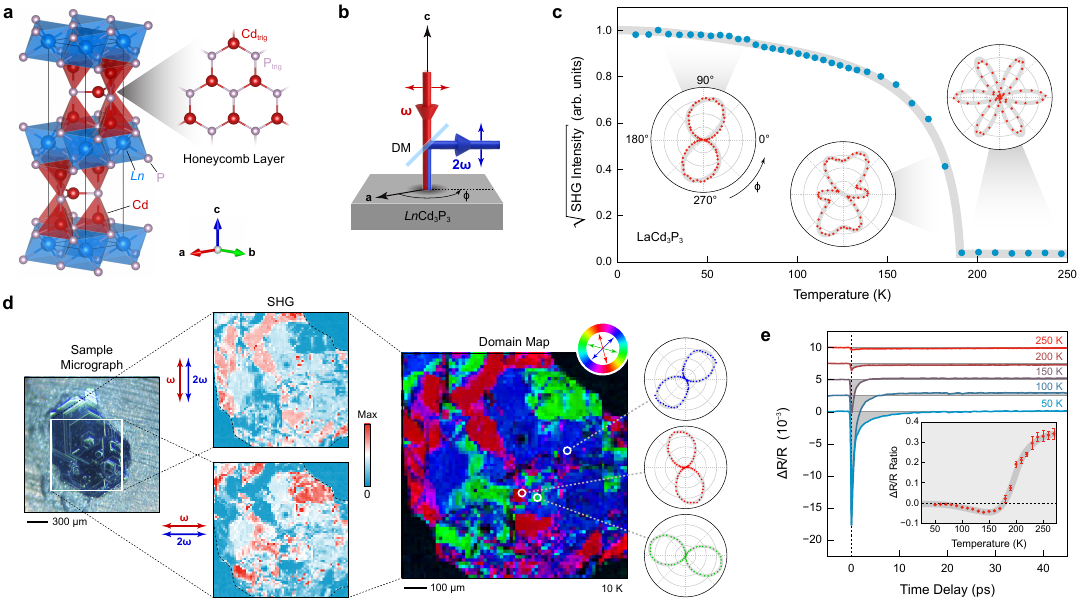}
\caption{\label{figure1} \textbf{Emergence of polar order and domain formation in LaCd$_3$P$_3$.} \textbf{a,}~Crystal structure of \textit{Ln}Cd$_3$P$_3$. Inset shows the two-dimensional honeycomb network of alternating Cd$_\textrm{trig}$ and P$_\textrm{trig}$ atoms that gives rise to the valence band states. \textbf{b,}~Geometry of RA-SHG measurements. $\phi$ denotes the angle of the beam polarization relative to the crystallographic $a$ axis. Small arrows indicate the linear polarization directions of the incident fundamental ($\omega$) and reflected second harmonic ($2\omega$) beams. DM, dichroic mirror. \textbf{c,}~Temperature dependence of the square root of SHG intensity for LaCd$_3$P$_3$. The onset of SHG below $T_c = 190$~K indicates an inversion-symmetry-breaking phase transition. Insets show representative RA-SHG patterns above, at, and below $T_c$, revealing spontaneous breaking of the in-plane rotational symmetry in the odd-parity phase. \textbf{d,}~SHG microscopy of LaCd$_3$P$_3$ at 10~K. Raster scans acquired with two perpendicular polarization geometries are combined to generate a domain map, revealing three distinct domain variants. Corresponding RA-SHG patterns demonstrate that the domains are related by 120$^\circ$ rotations. \textbf{e,}~Ultrafast transient reflectivity measurements of LaCd$_3$P$_3$ at selected temperatures. The $\Delta R/R$ ratio, defined in the text, exhibits a pronounced change across $T_c$, consistent with an electronic reconstruction accompanying the phase transition.}
\end{figure*}

\textit{Ln}Cd$_3$P$_3$ crystallizes in the hexagonal ScAl$_3$C$_3$ structure type (space group $P6_3/mmc$), consisting of alternating layers of octahedrally coordinated \textit{Ln}P$_6$, tetrahedrally coordinated CdP$_4$, and trigonal planar CdP$_3$ units~\cite{nientiedt1999}. The latter form two-dimensional honeycomb networks of alternating Cd$_\textrm{trig}$ and P$_\textrm{trig}$ atoms (Fig.~\ref{figure1}a). The crystallographic unit cell contains two honeycomb CdP layers related by inversion symmetry about the center of the cell, rendering the high-temperature structure globally centrosymmetric. \textit{Ln}Cd$_3$P$_3$ compounds are narrow-gap semiconductors whose low-energy electronic structure is derived predominantly from P$_\textrm{trig}$ $p_x$ and $p_y$ orbitals forming the valence band manifold~\cite{higuchi2016,feng2019,chamorro2023}. The degeneracy of the $p_x$ and $p_y$ orbitals provides a natural setting for interaction-driven symmetry breaking instabilities. Owing to the layered nature of the crystal structure, single crystals cleave readily along the basal plane perpendicular to the crystallographic $c$ axis.

Taking \textit{Ln} = La as a prototype system, we performed temperature-dependent RA-SHG measurements. In this technique, an ultrafast laser is incident normal to the sample surface and the reflected second harmonic intensity is measured as a function of the angle $\phi$ between the linear optical polarization and the crystallographic $a$ axis (Fig.~\ref{figure1}b). The second harmonic response arises from the material's nonlinear optical susceptibility tensor $\chi_{ijk}$ through $P_i(2\omega) = \chi_{ijk}E_j(\omega)E_k(\omega)$ and is therefore highly sensitive to crystal symmetry. Because electric-dipole SHG is forbidden in centrosymmetric systems, RA-SHG provides an especially sensitive probe of inversion symmetry breaking~\cite{harter2017,orenstein2021}. Figure~\ref{figure1}c shows the square root of the angle-integrated SHG intensity of LaCd$_3$P$_3$, proportional to $|\chi_{ijk}|$, as a function of temperature. At high temperatures, only a weak, temperature-independent surface contribution is observed, with six-lobed RA patterns consistent with the crystal surface symmetry. Upon cooling below 190~K, however, the SHG intensity exhibits a sharp order-parameter-like onset and eventually increases by nearly three orders of magnitude, heralding bulk inversion symmetry breaking. Simultaneously, the RA patterns evolve from six-fold to two-fold as the growing bulk response coherently mixes with the weak surface signal. At low temperatures, the response is dominated by a pronounced two-lobe pattern that starkly violates the six-fold rotational symmetry of the underlying lattice. These observations establish a second-order phase transition at $T_c = 190$~K that simultaneously breaks inversion and in-plane rotational symmetry.

To characterize the symmetry breaking, we performed SHG microscopy at 10~K by raster scanning the sample surface (Fig.~\ref{figure1}d). SHG images were acquired at two orthogonal polarization angles, $\phi_1$ and $\phi_2 = \phi_1 + 90^\circ$. Owing to the strong rotational anisotropy of the SHG response, the resulting intensities $I_1$ and $I_2$ vary according to the local orientation of the RA pattern. Heuristically, if the anisotropy follows an approximate angular dependence $I(\phi) \propto \cos^2(\phi - \phi_0)$, where $\phi_0$ denotes the orientation of the SHG lobes, then the two measurements yield intensities $I_1 \propto \cos^2(\phi_1 - \phi_0)$ and $I_2 \propto \sin^2(\phi_1 - \phi_0)$. Combining these intensities therefore allows the local orientation angle to be extracted according to $\theta = \tan^{-1}\sqrt{I_2/I_1} = \phi_1 - \phi_0$. We constructed a domain map by associating $\theta$ with hue and $I_1 + I_2$ with value. The resulting image reveals three dominant domain variants with characteristic sizes of $\sim$100~$\mu$m, corresponding to the red, green, and blue regions in Fig.~\ref{figure1}d. RA-SHG measurements within each domain show that the anisotropy axes differ by 120$^\circ$, consistent with a spontaneous $C_6 \rightarrow C_2$ rotational symmetry reduction. Whereas the high-temperature phase belongs to the centrosymmetric point group $6/mmm$ ($D_{6h}$), the highest-symmetry low-temperature group consistent with our measurements is the polar point group $mm2$ ($C_{2v}$). Quantitative fits based on $mm2$ symmetry accurately reproduce the measured RA-SHG patterns (gray curves in Fig.~\ref{figure1}c), confirming the emergence of an in-plane polar axis aligned along the crystallographic $a$ axis or one of its symmetry-equivalent directions.

To further probe the odd-parity phase transition, we performed ultrafast transient reflectivity measurements. In these pump-probe experiments, an ultrafast laser pulse excites the sample and the resulting relative change in reflectivity, $\Delta R/R$, is measured as a function of pump-probe time delay. Following photoexcitation, the reflectivity initially decreases ($\Delta R/R < 0$) before recovering on a few-picosecond timescale toward a longer-lived steady state whose sign depends on temperature (Fig.~\ref{figure1}e). To qualitatively track the evolution of the transient response while minimizing sensitivity to extrinsic measurement variation, we define a ratiometric quantity by averaging $\Delta R/R$ between 30 and 60~ps and normalizing by the magnitude of the initial minimum immediately following excitation (Extended Data Fig.~1). This ratio (inset of Fig.~\ref{figure1}e) exhibits a pronounced anomaly at $T_c$, including a substantial decrease and sign reversal upon cooling through the transition. Because the transient reflectivity on these timescales is governed primarily by the low-energy electronic structure and carrier relaxation processes, these changes provide evidence for a significant electronic reconstruction accompanying the onset of odd-parity order. In particular, the sign reversal indicates a substantial modification of carrier relaxation pathways and optical spectral weight across the transition. Notably, the effect begins slightly above $T_c$, where global symmetries remain unbroken but fluctuations of odd-parity order may already be significant. Together with the SHG results, these measurements suggest that the phase transition involves a substantial reorganization of the electronic states rather than a purely structural instability.

\subsection{Comparison across lanthanide series}

\begin{figure*}[t]
\includegraphics[width=6.4in]{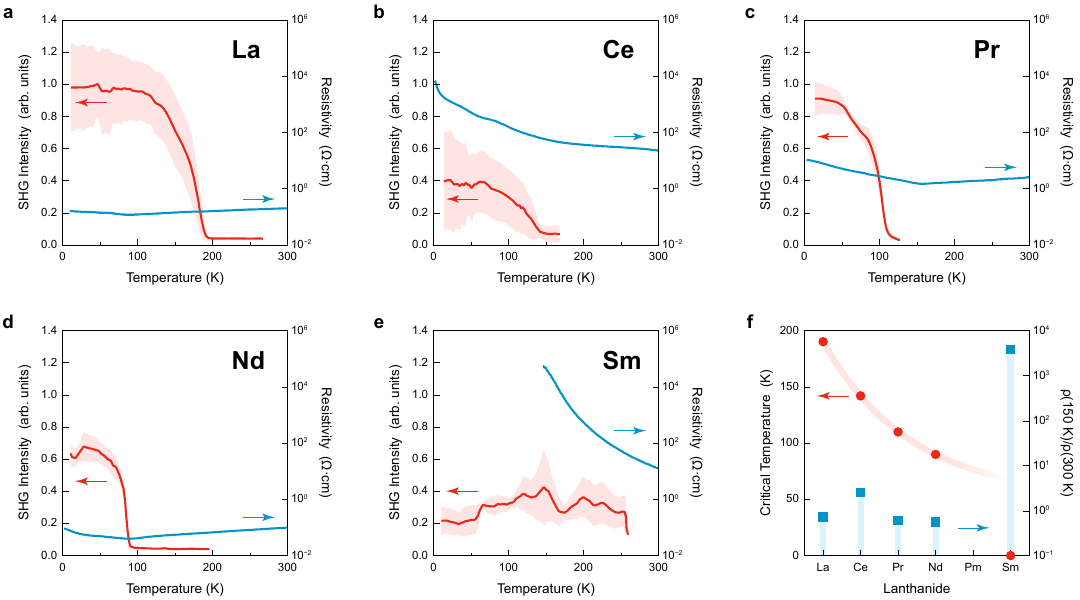}
\caption{\label{figure2} \textbf{Correlation between metallicity and odd-parity order across the \textit{Ln}Cd$_3$P$_3$ series.} \textbf{a-e,}~Temperature dependence of SHG intensity (red, left axis) and electrical resistivity (blue, right axis) for \textit{Ln} = La, Ce, Pr, Nd, and Sm. Pink shaded regions indicate the standard deviation of spot-to-spot variations in the SHG intensity obtained from spatial grid scans. La, Ce, Pr, and Nd remain metallic or weakly semiconducting and exhibit odd-parity phase transitions, whereas Sm is insulating and shows no detectable SHG onset. \textbf{f,}~Transition temperature $T_c$ (red, left axis) and resistivity ratio $\rho(150~\textrm{K})/\rho(300~\textrm{K})$ (blue, right axis) across the lanthanide series. $T_c$ smoothly decreases with increasing lanthanide atomic number. The absence of order in the Sm sample suggests that itinerant charge carriers promote the emergence of the odd-parity phase.}
\end{figure*}

Having established spontaneous odd-parity order in LaCd$_3$P$_3$, we next examined whether similar behavior occurs throughout the \textit{Ln}Cd$_3$P$_3$ family. Figure~\ref{figure2} summarizes temperature-dependent SHG measurements for \textit{Ln} = La, Ce, Pr, Nd, and Sm. With the notable exception of Sm, all compounds exhibit a clear onset of bulk SHG upon cooling, indicating the emergence of a globally noncentrosymmetric phase. The transition temperature evolves systematically across the series, decreasing monotonically from La to Nd with increasing atomic number (Fig.~\ref{figure2}f). This smooth trend suggests a common mechanism for odd-parity order throughout the series. Sm, however, deviates markedly from this behavior. Extrapolation of the lighter lanthanides yields an expected transition temperature near $70$~K, yet no global SHG onset is observed down to 10~K. All compounds also exhibit spatial variations in SHG intensity. The pink shaded regions in Fig.~\ref{figure2}a-e indicate the standard deviation obtained from spatial grid scans and quantify this intrinsic inhomogeneity.

A natural explanation for both the anomalous behavior of Sm and the spatial inhomogeneity is provided by the tendency of \textit{Ln}Cd$_3$P$_3$ to self-dope. Previous studies of flux-grown crystals reported metallic behavior with small hole concentrations~\cite{lee2019,chamorro2025,alvarado2025}. One proposed origin of this self-doping is excess interstitial P atoms above and below the CdP honeycomb layers, as directly observed by scanning transmission electron microscopy in the related compound GdZn$_3$P$_3$~\cite{guo2026}. Figure~\ref{figure2} compares electrical resistivity and SHG on the same temperature axis. La, Ce, Pr, and Nd compounds all exhibit weakly temperature-dependent resistivities characteristic of lightly doped semiconductors, with measured Hall carrier concentrations at the sub-percent level consistent with previous studies~\cite{lee2019,guo2026}. By contrast, Sm remains strongly insulating and is the only member of the series that does not exhibit a global odd-parity transition.

Anomalies in heat capacity, infrared spectra, and transport have been reported near the expected $T_c$ in metallic flux-grown single crystals but not in insulating polycrystalline samples~\cite{higuchi2016,lee2019,chamorro2023,jang2025}. Our results likewise demonstrate a close connection between the presence of itinerant holes and the emergence of odd-parity order: global inversion-symmetry breaking is observed only in samples with measurable electrical conduction. Self-doping also provides a natural source of disorder, as spatial variations in defect concentration generate corresponding variations in local carrier density. The Ce compound illustrates this interplay particularly clearly. Although it undergoes an odd-parity transition, it exhibits weakly insulating transport, unusually large SHG variations, and a substantially more disordered domain texture than LaCd$_3$P$_3$ (Extended Data Fig.~2). These observations suggest that Ce may lie near a boundary between globally connected and disconnected conducting regions, where disorder strongly influences both transport and symmetry breaking. While the microscopic origin of the self-doping remains to be fully understood, the combined SHG and transport data consistently indicate that a small density of mobile holes plays a key role in stabilizing the odd-parity phase.

\subsection{Theoretical model of odd-parity order}

\begin{figure*}[t]
\includegraphics[width=6.4in]{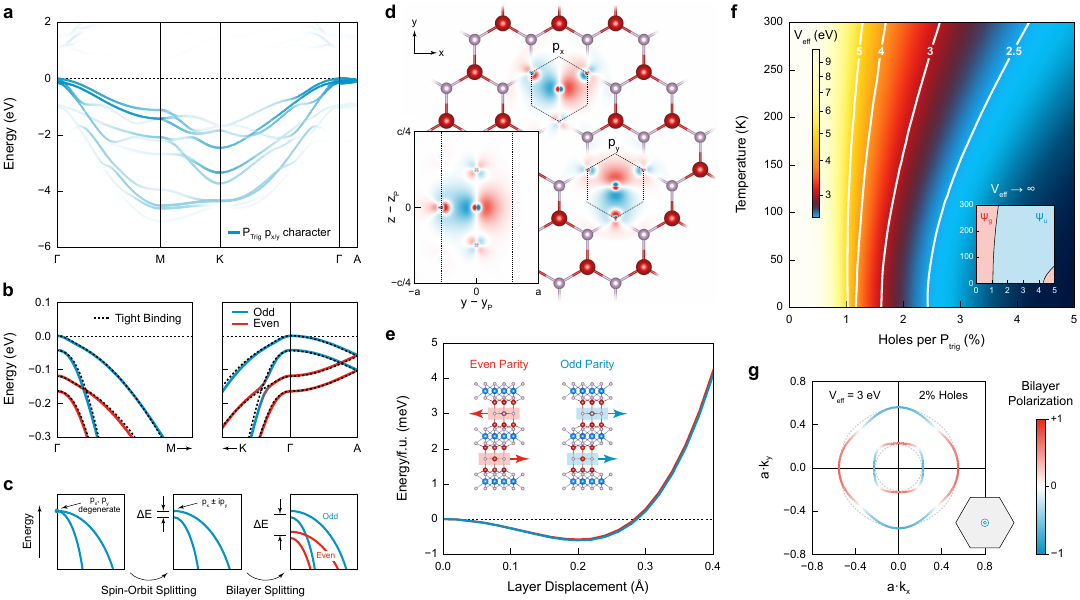}
\caption{\label{figure3} \textbf{Electronic mechanism for polar order in LaCd$_3$P$_3$.} \textbf{a,}~DFT band structure of LaCd$_3$P$_3$. The valence band maximum at $\Gamma$ is dominated by P$_\textrm{trig}$ $p_x$ and $p_y$ character, as indicated by the band intensity. \textbf{b,}~Expanded view of the valence band dispersion near $\Gamma$. The overlaid dashed curves show the corresponding tight-binding model, demonstrating excellent agreement with DFT. \textbf{c,}~Schematic illustration of band splittings at $\Gamma$. In the absence of spin-orbit coupling, the $p_x$ and $p_y$ bands are degenerate. Spin-orbit coupling lifts this degeneracy by $\Delta E = 44$~meV, while interlayer hybridization produces an additional bilayer splitting of $\Delta E = 120$~meV. \textbf{d,}~Localized Wannier orbitals with predominant P$_\textrm{trig}$ $p_x$ and $p_y$ character used to construct the tight-binding model. \textbf{e,}~Total DFT energy as a function of even- and odd-parity structural distortions corresponding to relative sliding of the CdP honeycomb layers with respect to the rest of the unit cell. Both distortions exhibit comparable energetics, with a local minimum at a layer displacement of 0.2~\AA. \textbf{f,}~Doping-temperature phase diagram showing the critical interaction strength $V_\mathrm{eff}$ required to stabilize $\psi_u$ order. Inset shows the relative stability of $\psi_g$ and $\psi_u$ when the interaction strength is sufficiently large. \textbf{g,}~Tight-binding Fermi surface reconstruction near $\Gamma$ for $V_\mathrm{eff} = 3$~eV and 2\% holes per P$_\textrm{trig}$. Dashed lines denote the unreconstructed Fermi surface. The reconstructed Fermi surface acquires a bilayer polarization absent in the centrosymmetric phase, directly reflecting inversion symmetry breaking. The inset shows the relative size of the Fermi sheets with respect to the entire first Brillouin zone.}
\end{figure*}

The experimental results establish spontaneous odd-parity order throughout the \textit{Ln}Cd$_3$P$_3$ family. Multiple observations suggest that this instability is electronic rather than structural in origin. First, ultrafast transient reflectivity reveals a substantial electronic reconstruction at the transition. Second, odd-parity order appears only in samples with mobile hole carriers, whereas the most insulating member of the series remains globally centrosymmetric. These findings point to an interaction-driven instability of the low-energy electronic states. In a Fermi liquid, such symmetry breaking is naturally described within the framework of a Pomeranchuk instability, in which interactions drive a Fermi surface distortion and reduce its symmetry~\cite{pomeranchuk1959}. The appearance of such an instability in \textit{Ln}Cd$_3$P$_3$ is, however, highly unusual because the carrier densities inferred from Hall measurements are extremely small, implying a correspondingly small Fermi surface and density of states. Since interaction-driven instabilities are generally strongest at large densities of states, the emergence of odd-parity order is not obvious. To address this puzzle, we developed a DFT-derived tight-binding model and show within mean-field theory that relatively modest interactions can stabilize odd-parity electronic order even at low carrier densities.

Figure~\ref{figure3}a shows the DFT valence band structure of LaCd$_3$P$_3$. The projected P$_\mathrm{trig}$ $p_x$ and $p_y$ orbital character demonstrates that the valence band maximum at $\Gamma$, where doped holes reside, derives almost entirely from these orbitals. A magnified view near $\Gamma$ is shown in Fig.~\ref{figure3}b. Four closely spaced bands are present, reflecting two distinct energy splittings illustrated schematically in Fig.~\ref{figure3}c. The first arises from spin-orbit coupling through hybridization with Cd$_\mathrm{trig}$ $d_{xy}$ and $d_{x^2-y^2}$ orbitals, lifting the $p_x$-$p_y$ degeneracy and producing $(p_x \pm i p_y)$-like states separated by $\Delta E = 44$~meV. Each of these states is further split by a larger bilayer hybridization of $\Delta E = 120$~meV between the two inversion-related CdP honeycomb layers in the unit cell, yielding states of definite inversion parity. This low-energy structure is captured by a minimal four-band tight-binding model comprising $p_x$ and $p_y$ orbitals on the two P$_\mathrm{trig}$ sites, one in each honeycomb layer. The corresponding DFT-derived Wannier orbitals are shown in Fig.~\ref{figure3}d. Remarkably, nearest-neighbor in-plane $\sigma$ and $\pi$ hopping together with interlayer $\pi$ hopping reproduce the DFT bands with excellent accuracy. As shown by the dashed curves in Fig.~\ref{figure3}b, this simple model captures the essential physics of the valence band manifold.

Before considering interaction-driven instabilities, it is important to assess whether the observed polar phase could arise from a conventional structural distortion. DFT calculations indeed reveal a weak lattice instability associated with sliding of the CdP honeycomb layers relative to the rest of the unit cell. As shown in Fig.~\ref{figure3}e, however, odd-parity (in-phase) and even-parity (out-of-phase) distortions are nearly degenerate, indicating the lattice alone exhibits little preference for a globally inversion-breaking distortion. Moreover, the associated energy landscape is extremely shallow, with a maximum energy lowering of only 0.6~meV (7~K) per formula unit, more than an order of magnitude smaller than the transition temperature. Although this soft mode may enhance an electronically-driven instability through electron-phonon coupling, it appears insufficient, by itself, to account for the robust odd-parity order observed near 190~K. We therefore do not include such coupling in our minimal model. The \textit{Ln}Cd$_3$P$_3$ family also exhibits Cd-P bond frustration within the honeycomb layers, producing short-range bond length disorder and an intrinsically inhomogeneous local lattice environment~\cite{chamorro2025,alvarado2025}. Because these distortions remain correlated over only a few unit cells, preserve the average crystal symmetry, and show no x-ray scattering anomalies at $T_c$, they neither generate the observed long-range inversion and rotational symmetry breaking nor couple linearly to the electronic order parameters considered below. We therefore neglect them as well.

We introduce electronic interactions by augmenting the tight-binding model with local Hubbard-like repulsion between the orbitals. The interaction Hamiltonian is given by
$${\mathcal{H}_\mathrm{int} = U \sum_{i,l} \left( \hat n_{x \uparrow l}^i \hat n_{x \downarrow l}^i + \hat n_{y \uparrow l}^i \hat n_{y \downarrow l}^i \right) + V \sum_{i,l} \left( \hat n_{x \uparrow l}^i + \hat n_{x \downarrow l}^i \right)\left( \hat n_{y \uparrow l}^i + \hat n_{y \downarrow l}^i \right),}$$
where $\hat n_{o \sigma l}^i$ denotes the density operator for orbital $o \in \{x,y\}$ and spin $\sigma \in \{\uparrow,\downarrow\}$ on honeycomb layer $l \in \{1,2\}$ of lattice site $i$, while $U$ and $V$ represent intra- and inter-orbital Coulomb repulsion energies, respectively. We treat these interactions within a mean-field approximation and focus on the orbital polarization order parameters
\begin{align*}
\psi_g &= \frac{1}{2} \sum_\sigma \left( \left\langle \hat n_{x \sigma 1} \right\rangle - \left\langle \hat n_{y \sigma 1} \right\rangle + \left\langle \hat n_{x \sigma 2} \right\rangle - \left\langle \hat n_{y \sigma 2} \right\rangle \right) \\
\psi_u &= \frac{1}{2} \sum_\sigma \left( \left\langle \hat n_{x \sigma 1} \right\rangle - \left\langle \hat n_{y \sigma 1} \right\rangle - \left\langle \hat n_{x \sigma 2} \right\rangle + \left\langle \hat n_{y \sigma 2} \right\rangle \right),
\end{align*}
which capture spontaneous orbital polarization within the $p_x$-$p_y$ manifold. Both order parameters correspond to an imbalance in the occupations of the $p_x$ and $p_y$ orbitals and therefore break the in-plane rotational symmetry of the honeycomb lattice. The distinction lies in how this imbalance is distributed between the two inversion-related honeycomb layers. For $\psi_g$, the orbital polarization has the same sign on both layers, yielding an even-parity state that preserves inversion symmetry. In contrast, $\psi_u$ corresponds to opposite orbital polarizations on the two layers. Because the layers are related by inversion, this orbital polarization is odd under inversion and therefore produces a state that simultaneously breaks rotational and inversion symmetry.

To determine the leading interaction-driven instability, we analyze the mean-field free energy as a function of $\psi_g$ and $\psi_u$. For each temperature and hole concentration, we evaluate the second derivatives of the grand canonical free energy with respect to the two order parameters. The point at which the corresponding curvature changes sign defines the critical interaction strength for ordering. Within this analysis, the intra- and inter-orbital interactions enter only through the effective parameter $V_\mathrm{eff} = V/2 - U/4$, which governs the overall tendency toward orbital polarization. The resulting phase diagram is shown in Fig.~\ref{figure3}f. The color scale and contours indicate the critical $V_\mathrm{eff}$ required to stabilize $\psi_u$, while the inset compares the stability of the even- and odd-parity phases above this threshold. The odd-parity state is favored over a broad range of temperatures and dopings. Remarkably, the required interaction strength remains modest even at percent-level hole doping, demonstrating that dilute carriers can drive inversion-breaking electronic order in this model. Although the ordered state is not fundamentally spin driven, spin-orbit coupling plays a crucial role by generating the band splittings that enhance the instability; in its absence, unrealistically large interactions are required to obtain ordering. These results provide a natural explanation for the experimentally observed polar phase despite the exceptionally low carrier densities in \textit{Ln}Cd$_3$P$_3$.

The order parameters $\psi_g$ and $\psi_u$ are associated with Pomeranchuk instabilities of the low-energy Fermi surface. Figure~\ref{figure3}g illustrates the resulting reconstruction for $V_\mathrm{eff} = 3$~eV and a hole concentration of 2\% per P$_\mathrm{trig}$, for which the model predicts $T_c \approx 160$~K. In the parent state, the Fermi surface consists of two nearly circular sheets (gray dashed lines) that preserve the six-fold rotational symmetry of the crystal. In the ordered phase, both sheets distort spontaneously, producing the characteristic rotational symmetry breaking of a Pomeranchuk instability. More strikingly, the odd-parity state develops a momentum-dependent bilayer polarization. Whereas the high-temperature electronic states are equally distributed between the two inversion-related honeycomb layers, the $\psi_u$ phase exhibits a layer polarization approaching 70\% along the $k_x$ and $k_y$ directions. This polarization corresponds to an occupation imbalance between the layers and therefore changes sign under inversion, which exchanges them. Because the polarization texture remains symmetric under $\mathbf{k} \rightarrow -\mathbf{k}$, it represents a fundamentally odd-parity electronic structure. No analogous texture appears for $\psi_g$. The odd-parity phase is therefore distinguished not only by a Fermi surface distortion but also by the emergence of a bilayer polarization texture, making $\psi_u$ an unconventional Pomeranchuk instability in which rotational and inversion symmetry are broken simultaneously.

\section{Discussion}

$V_\mathrm{eff}$ is the key parameter controlling the onset of odd-parity order. Although it enters our theory phenomenologically, its magnitude can be estimated from Coulomb matrix elements of the DFT-derived Wannier orbitals. Neglecting screening, we obtain $V_\mathrm{eff} = 2.0$~eV for strictly on-site interactions and $V_\mathrm{eff} = 4.2$~eV when nearest-neighbor repulsion is included. The latter contribution is substantial because the Wannier orbitals are not tightly localized on the P$_\mathrm{trig}$ sites but have significant weight on neighboring Cd$_\mathrm{trig}$ atoms (Fig.~\ref{figure3}d), leading to appreciable overlap between adjacent orbitals. These estimates place the required interaction strength within a physically realistic range. The precise value of $V_\mathrm{eff}$ will inevitably be modified by screening, whose treatment in layered semiconductors remains an active area of investigation~\cite{tuan2018,galiautdinov2019}. Nevertheless, screening is expected to be weak in \textit{Ln}Cd$_3$P$_3$, as the honeycomb-derived valence bands are relatively isolated from other electronic states and the carrier density is low. Moreover, our model neglects additional mechanisms that could further stabilize the ordered phase, including long-range dipolar interactions generated by the odd-parity state and coupling to the soft lattice mode discussed above. Indeed, a recent Raman spectroscopy work reports evidence of phonon softening in PrCd$_3$P$_3$ consistent with our findings~\cite{davis2026}. Taken together, these considerations suggest that the electronic interaction strengths required to stabilize odd-parity order are physically reasonable.

Another question concerns the hole concentration in as-grown crystals. Hall measurements on \textit{Ln}Cd$_3$P$_3$ and related materials typically report carrier densities on the order of 0.1\% holes per P$_\mathrm{trig}$~\cite{lee2019,guo2026}, smaller than the concentrations for which our model most readily stabilizes odd-parity order. However, several observations suggest that Hall measurements may substantially underestimate the total hole density. Most notably, scanning transmission electron microscopy on the closely related compound GdZn$_3$P$_3$ revealed an excess P concentration of approximately 0.04 per formula unit~\cite{guo2026}, more than an order of magnitude larger than the reported Hall carrier density. Although the precise relationship between these defects and the mobile carrier concentration remains unclear, this discrepancy suggests that transport measurements may not fully capture the electronic doping. Electrical transport is further complicated by thin crystal morphology and large contact resistances~\cite{guo2026}. In addition, our SHG measurements reveal pronounced spatial inhomogeneity, indicating that self-doping likely introduces substantial disorder and corresponding variations in local carrier density. Under such conditions, transport may be dominated by percolative conduction pathways and probe only a subset of the doped regions. We therefore regard the Hall density as a lower bound on the total hole concentration. Taken together, the structural, transport, and optical data are consistent with an actual hole concentration at the percent level, even if only a fraction of the carriers contribute to long-range charge transport.

In summary, we uncovered spontaneous symmetry breaking in the \textit{Ln}Cd$_3$P$_3$ material family. SHG measurements revealed an ordered phase that simultaneously breaks inversion and rotational symmetry and develops an in-plane polar axis. The transition occurs throughout the lanthanide series provided a small density of hole carriers is present, suggesting an unconventional mechanism distinct from those typically associated with ferroelectrics or strongly correlated systems. Guided by a DFT-derived tight-binding model, we showed that modest electronic interactions can stabilize an odd-parity instability in which a rotationally distorted Fermi surface acquires a momentum-dependent bilayer polarization texture. The resulting state is therefore more naturally viewed as an odd-parity analogue of a Pomeranchuk instability than as a conventional ferroelectric transition. Its stabilization at low carrier density is particularly remarkable, as interaction-driven Fermi surface instabilities are typically associated with large densities of states or proximity to Van Hove singularities, neither of which is present in \textit{Ln}Cd$_3$P$_3$. Instead, the combination of nearly degenerate orbitals, spin-orbit coupling, inversion-related honeycomb bilayers, and moderate Coulomb interactions appears sufficient to drive parity-breaking order near the semiconductor limit. These results establish \textit{Ln}Cd$_3$P$_3$ as a rare example of a dilute-carrier electronic system exhibiting spontaneous inversion symmetry breaking and identify honeycomb bilayer semiconductors as a promising platform for interaction-driven odd-parity phases.

\section{Methods}

\subsection{Sample synthesis and characterization}
Single crystals of \textit{Ln}Cd$_3$P$_3$ were prepared using a molten salt flux method~\cite{nientiedt1999}. Samples were synthesized using a two-step process. In the first step, the binary precursor \textit{Ln}Cd$_2$ was synthesized. For this purpose, stoichiometric amounts of fresh \textit{Ln} filings were combined with Cd shot (Alfa Aesar, 99.999\% metals basis, 3-6~mm) in a glove box. The mixture was transferred into an alumina Canfield crucible and sealed in a silica ampoule under Ar gas at a pressure of 252~Torr. The mixture was reacted at 800$^\circ$C for 8~hr at a slow heating rate of 50$^\circ$C/hr to avoid bursting of the ampoule by the high vapor pressure of Cd. In the second step, the binary was reacted with Cd shot and red P powder (BeanTown Chemical, 99.9999\%) in a stoichiometric ratio to make 0.5~g of the solute mixture. A 2\% weight excess of P was used to account for weight loss when P sticks to the weight boat. To this solute mixture, 2~g of a 1:1 atomic mixture of NaCl and KCl was added. The salt flux was preheated at a temperature of 700$^\circ$C overnight to drive off any absorbed moisture. This mixture was transfered to a Canfield crucible and sealed in a silica ampoule of thickness 2~mm under Ar gas at a pressure of 252~Torr. Crystal growth followed the heating profile described in Ref.~\citenum{alvarado2025}, after which the ampoule was broken under ambient atmosphere and the grown crystals were extracted from the salt flux by dissolution in deionized water. Thin, plate-like crystals of lateral dimensions 2-3~mm and thicknesses varying from 20-100~$\mu$m were produced. The electrical resistivity of samples was measured using the Electrical Transport Option (ETO) installed on a Quantum Design DynaCool system. Four-probe electrical contacts were made on the plate-like crystals with 0.05~mm diameter gold wires and silver epoxy. Before the measurement, the silver epoxy was cured in an oven for 1~hr for better contact with the crystal surface.

\subsection{Optical experiments}
Optical measurements were performed on the (001) surfaces of as-grown single crystals mounted in an optical cryostat. Ultrafast laser pulses (40~fs, 50~kHz repetition rate) were generated with a typical fluence of $\sim$1~mJ/cm$^{2}$ and center wavelengths of 800~nm and 1515~nm for the probe and pump beams, respectively. For static SHG measurements, the probe beam was incident normal to the sample surface (Fig.~\ref{figure1}b). The reflected second harmonic signal was spectrally filtered and detected with a silicon photomultiplier and lock-in amplifier. RA-SHG patterns were acquired by rotating the incident polarization and detector simultaneously about the surface normal while keeping the sample fixed, which is equivalent to rotating the crystal. Measurements were performed in a parallel-polarization geometry, but crossed-polarization measurements yielded similar results. To characterize spatial variations in the SHG response, measurements were repeated over grids of surface locations with a pitch of 50~$\mu$m. Depending on the available crystal area, grid sizes ranged from $2 \times 2$ to $5 \times 5$ points. For ultrafast pump-probe measurements, the probe beam was incident at an angle of 10$^\circ$ while the pump remained normally incident. The pump-probe delay was controlled with a mechanical delay stage, and the transient reflectivity $\Delta R/R$ was recorded as a function of delay time.

\subsection{SHG symmetry analysis and fitting}
The nonlinear optical susceptibility tensor $\chi_{ijk}$ is constrained by crystal symmetry and therefore provides a direct probe of symmetry breaking. The high-temperature phase of \textit{Ln}Cd$_3$P$_3$ belongs to the centrosymmetric space group $P6_3/mmc$, for which all bulk electric-dipole contributions to the third-rank susceptibility tensor vanish identically. The crystal surface, however, necessarily breaks inversion symmetry. The weak six-fold RA-SHG pattern observed above $T_c$ is therefore attributed to a surface response with point group symmetry $3m$, obtained after the surface breaks the out-of-plane mirror, screw axis, and glide plane symmetries of the bulk crystal. At normal incidence, this contribution takes the form $I_\mathrm{surf}(\phi) = A^2\sin^2(3\phi)$, with $A = \chi^\mathrm{(surf)}_{xxy}$. The abrupt enhancement of SHG below $T_c$ signals the onset of bulk electric-dipole SHG and therefore spontaneous inversion-symmetry breaking. Among the subgroups of the parent point group $6/mmm$, the observed two-fold anisotropy at normal incidence is consistent with point group $mm2$ and its subgroups, with the two-fold axis lying in the (001) plane. For $mm2$, the bulk contribution is $I_\mathrm{bulk}(\phi) = \left(B_1\cos^2(\phi)\sin(\phi) + B_2\sin^3(\phi)\right)^2$, where $B_1 = 2\chi^\mathrm{(bulk)}_{yyz} + \chi^\mathrm{(bulk)}_{zyy}$ and $B_2 = \chi^\mathrm{(bulk)}_{xzz}$. Near $T_c$, surface and bulk SHG contributions are comparable and interfere coherently. The resulting intensity is therefore $I_\mathrm{tot}(\phi) = \left|B_1\cos^2(\phi)\sin(\phi) + B_2\sin^3(\phi) + Ae^{i\gamma}\sin(3\phi)\right|^2$, where $\gamma$ accounts for the relative phase between the surface and bulk responses. This expression was used to fit all RA-SHG patterns presented in Fig.~\ref{figure1}.

\subsection{Density functional theory calculations}
Band structure calculations were performed using the Vienna Ab initio Simulation Package (\textsc{vasp})~\cite{kresse1994,kresse1996a,kresse1996b,kresse1999} with the Perdew-Burke-Ernzerhof (PBE) parameterization of the generalized gradient approximation~\cite{perdew1996} and the supplied projector augmented wave pseudopotentials. Fully-relaxed crystal structure parameters ($a = 4.2972$~\AA, $c = 21.0463$~\AA) were used together with a plane-wave cutoff of 500~eV, an $8 \times 8 \times 4$ $\Gamma$-centered k-point mesh, and D3 van der Waals corrections~\cite{grimme2010}. Spin-orbit coupling was included self-consistently. Wannier orbitals were constructed using the full-potential code WIEN2k~\cite{blaha2018,blaha2020}. These calculations used $R_{\mathrm{MT}}K_{\mathrm{max}} = 8.0$ and an $8 \times 8 \times 4$ $\Gamma$-centered k-point mesh.

\subsection{Tight-binding model}
To describe the low-energy electronic structure of \textit{Ln}Cd$_3$P$_3$, we construct a minimal four-band tight-binding model consisting of $p_x$ and $p_y$ orbitals on the two P$_\mathrm{trig}$ sites in the crystallographic unit cell, one belonging to each inversion-related honeycomb layer. The Hamiltonian includes nearest-neighbor in-plane $\sigma$ and $\pi$ hopping ($t^\parallel_\sigma$ and $t^\parallel_\pi$), interlayer $\pi$ hopping ($t^\perp_\pi$), and spin-orbit coupling. The hopping amplitudes are evaluated using standard Slater-Koster matrix elements~\cite{slater1954}, giving
$${\mathcal{H}_0(\mathbf{k}) = \left[\begin{array}{cccc}
t^\parallel_{xx} & t^\parallel_{xy} + i\lambda & t^\perp & 0	\\
t^\parallel_{xy} - i\lambda & t^\parallel_{yy} & 0 & t^\perp	\\
t^\perp & 0 & t^\parallel_{xx} & t^\parallel_{xy} + i\lambda	\\
0 & t^\perp & t^\parallel_{xy} - i\lambda & t^\parallel_{yy}
\end{array}\right]}$$
with
\begin{align*}
t^\parallel_{xx} &= t^\parallel_\sigma\left[ 2\cos(k_a) + 1/2\cos(k_b) + 1/2\cos(k_a + k_b) \right] + t^\parallel_\pi\left[ 3/2\cos(k_b) + 3/2\cos(k_a + k_b) \right] \\
t^\parallel_{yy} &= t^\parallel_\sigma\left[ 3/2\cos(k_b) + 3/2\cos(k_a + k_b) \right] + t^\parallel_\pi\left[ 2\cos(k_a) + 1/2\cos(k_b) + 1/2\cos(k_a + k_b) \right] \\
t^\parallel_{xy} &= t^\parallel_\sigma\left[ \sqrt{3}/2\cos(k_a + k_b) - \sqrt{3}/2\cos(k_b) \right] + t^\parallel_\pi\left[ \sqrt{3}/2\cos(k_b) - \sqrt{3}/2\cos(k_a + k_b) \right] \\
t^\perp &= t^\perp_\pi\cos(k_c/2),
\end{align*}
where $k_a$, $k_b$, and $k_c$ are the projections of $\mathbf{k}$ onto the three lattice vectors. Spin-orbit coupling is parameterized by $\lambda$, producing a splitting $\Delta E = 2\lambda$, while interlayer hybridization generates a bilayer splitting $\Delta E = 2t^\perp_\pi$. Diagonalization of $\mathcal{H}_0(\mathbf{k})$ yields four band energies $\xi_{n\mathbf{k}}$ ($n = 1, 2, 3, 4$). The DFT valence band dispersion is accurately reproduced using $t^\parallel_\sigma = 1.7$~eV, $t^\parallel_\pi = -0.3$~eV, $t^\perp_\pi = 0.06$~eV, and $\lambda = 0.022$~eV (Fig.~\ref{figure3}b).

\subsection{Mean-field theory}
The interaction Hamiltonian $\mathcal{H}_\mathrm{int}$ is treated within mean-field theory by considering the even- and odd-parity order parameters $\psi_g$ and $\psi_u$ separately. The resulting mean-field interaction takes the form
$${\mathcal{H}^\mathrm{MF}_\mathrm{int} = V_\mathrm{eff} \left[\begin{array}{cccc}
-\psi & 0 & 0 & 0	\\
0 & +\psi & 0 & 0	\\
0 & 0 & \pm\psi & 0	\\
0 & 0 & 0 & \mp\psi
\end{array}\right],}$$
where $\psi = \psi_u$ for the upper signs and $\psi = \psi_g$ for the lower signs. Adding $\mathcal{H}^\mathrm{MF}_\mathrm{int}$ to $\mathcal{H}_0(\mathbf{k})$ and diagonalizing yields the interacting eigenvalues $\tilde{\xi}_{n\mathbf{k}}$. The grand canonical free energy per site is
$${\Omega = -\frac{k_BT}{N}\sum_{n,\mathbf{k}}\ln\left[ 1 + e^{-(\tilde{\xi}_{n\mathbf{k}} - \mu)/k_BT} \right] + f(\psi),}$$
where $N$ is the total number of lattice sites, the chemical potential $\mu$ is determined self-consistently for each temperature and hole concentration, and $f(\psi) = V_\mathrm{eff}\psi^2$ is the mean-field energy offset. To identify the onset of ordering, we evaluate the curvature of the free energy at $\psi = 0$,
$${\left.\frac{\partial^2\Omega}{\partial\psi^2}\right|_{\psi = 0} = \frac{1}{N}\sum_{n,\mathbf{k}} \left[ n_F(\xi_{n\mathbf{k}} - \mu)\frac{\partial^2\tilde{\xi}_{n\mathbf{k}}}{\partial\psi^2} - \frac{1}{4k_BT\cosh^2[(\xi_{n\mathbf{k}} - \mu)/2k_BT]}\left(\frac{\partial\tilde{\xi}_{n\mathbf{k}}}{\partial\psi}\right)^2 \right] + 2V_\mathrm{eff},}$$
where $n_F$ is the Fermi-Dirac distribution and all derivatives of $\tilde{\xi}_{n\mathbf{k}}$ are evaluated at $\psi = 0$. The phase boundary is defined by the point at which this curvature changes sign. Solving the condition $\partial^2\Omega/\partial\psi^2 = 0$ for $V_\mathrm{eff}$ yields the critical interaction strength shown in Fig.~\ref{figure3}f. Momentum-space sums were evaluated on a $2000 \times 2000 \times 400$ k-point grid. We find that spin-orbit coupling is essential for the instability: setting $\lambda = 0$ eliminates both $\psi_u$ and $\psi_g$ ordering within the physically reasonable parameter range.

\section{Data Availability}

All data that support the findings of this study are presented in the main text and the extended data figures. This data is available from the corresponding author upon reasonable request.

\section{Acknowledgments}

This work was supported by the National Science Foundation (NSF) under Award No.~DMR-2140786. S.D.W. and D.R. were supported by the U.S. Department of Energy (DOE), Office of Basic Energy Sciences, Division of Materials Sciences and Engineering under Grant No.~DE-SC0017752. Use was made of shared facilities supported by the Materials Research Science and Engineering Center (MRSEC) Program of the NSF under Award No.~DMR-2308708, including computational facilities purchased with funds from the NSF (CNS-1725797) and administered by the Center for Scientific Computing (CSC). The CSC is supported by the California NanoSystems Institute.

\section{Author Contributions}

J.T. (Jack Tregidga), under the supervision of J.W.H., conducted the optical measurements, analyzed the data, and performed the DFT calculations. D.R., J.H., and J.T. (Josiah Turner), under the supervision of S.D.W., synthesized and characterized the samples. J.T. (Jack Tregidga) and J.W.H. wrote the manuscript, with input from all authors.

\section{Competing Interests}

The authors declare no competing interests.

\clearpage

\begin{figure*}[t]
\includegraphics{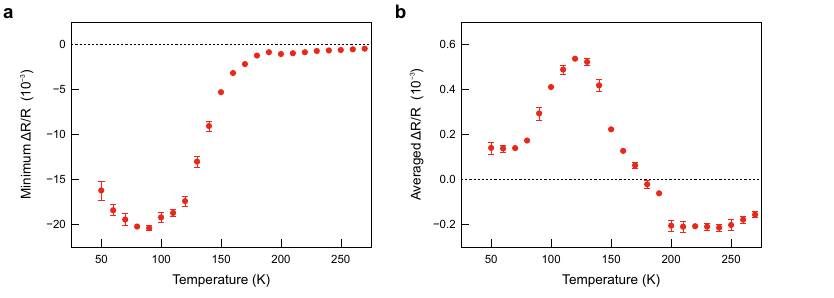}
\par\justifying\noindent{\textbf{Extended Data Fig.~1 \textbar~Transient reflectivity data for LaCd$_3$P$_3$.} \textbf{a,}~Magnitude of the initial minimum in $\Delta R/R$ immediately following excitation by the pump pulse. \textbf{b,}~Average value of $\Delta R/R$ between 30 and 60~ps, capturing the offset at longer time scales. The ratio of these two quantities is displayed in the inset of Fig.~\ref{figure1}e.}
\end{figure*}

\begin{figure*}[t]
\includegraphics{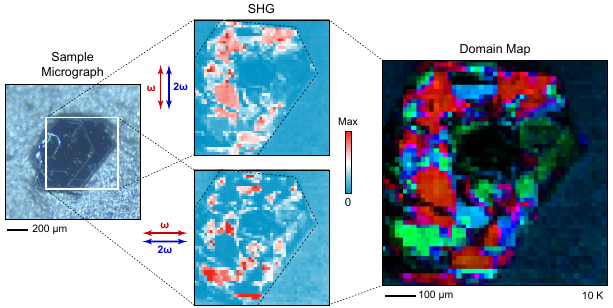}
\par\justifying\noindent{\textbf{Extended Data Fig.~2 \textbar~SHG microscopy of CeCd$_3$P$_3$.} Raster scans acquired with two perpendicular polarization geometries are combined to generate a domain map. Smaller domains and more disorder are observed in the Ce compound as compared to La. A large region near the center of the sample shows no appreciable SHG, which may arise from inhomogeneous self-doping.}
\end{figure*}

\end{document}